\documentclass[twocolumn,showpacs,preprintnumbers,amsmath,amssymb,nofootinbib]{revtex4}

\usepackage{graphicx}
\usepackage{dcolumn}
\usepackage{bm}
\usepackage{epsf}

\begin{document}


\title{A model-independent test for scale-dependent non-Gaussianities in the cosmic microwave background}

\author{C. R\"ath$^1$, G. E. Morfill$^1$, G. Rossmanith$^1$, A. J. Banday$^2$, K. M. G\'{o}rski$^{3,4}$}
\affiliation{$^1$Max-Planck-Institut f\"ur extraterrestrische Physik, Giessenbachstr.1, 85748 Garching, Germany\\
 $^2$Centre d'Etude Spatiale des Rayonnements, 9, Av du Colonel Roche, 31028 Toulouse, France\\
 $^3$Jet Propulsion Laboratory, California Institute of Technology, Pasadena, CA 91109, USA\\
  $^4$Warsaw University Observatory, Aleje Ujazdowskie 4, 00 - 478 Warszawa, Poland
}

\date{\today}

\begin{abstract}

We present a model-independent method to test for scale-dependent 
non-Gaussianities in combination with scaling indices
as test statistics. Therefore, surrogate data sets are generated, 
in which the power spectrum of the original data is preserved, 
while the higher order correlations are partly randomised by  
applying a scale-dependent shuffling procedure to the Fourier phases.
We apply this method to the WMAP data of the 
cosmic microwave background (CMB) and find signatures
for non-Gaussianities on large scales. Further tests are required to elucidate
the origin of the detected anomalies.

\end{abstract}


\maketitle


Inflationary models of the very early universe have proved to 
be in very good agreement with the observations of the 
linear correlations of the cosmic microwave background (CMB). 
While the simplest, single field, slow-roll inflation  
\cite{Guth81,Linde82,Albrecht82} predicts
that the temperature fluctuations of the CMB 
correspond to a (nearly) Gaussian, homogeneous and isotropic random field,
more complex  models may  give rise
to non-Gaussianity  \cite{Linde97,Peebles97,Bernardeau02,Acquaviva03}.
Models in which the Lagrangian is a general function of the 
inflaton and powers of its first derivative \cite{Armendariz99,Garriga99} 
can lead to scale-dependent
non-Gaussianities, if the sound speed varies during inflation. 
Similarily, string theory models that give rise to large non-Gaussianity 
have a natural scale dependence \cite{Loverde08a}. 
If the scale dependence of non-Gaussian signatures plays an important
role in theory,  the conventional  (global) parametrisation of 
non-Gaussianity via $f_{NL}$ is no longer  sufficient to describe the level
of non-Gaussianity and to discriminate 
between different models. $f_{NL}$ must at least become scale dependent - if
this parametrisation is sufficient at all. 
But first of all such scale-dependent signatures
have to be identified.\\
Possible deviations from Gaussianity  have been investigated in 
studies based on e.g. the WMAP data of the CMB 
(see \cite{Komatsu09a} and references therein) and claims for the detection of 
non-Gaussianities and other anomalies (see e.g. 
\cite{Park04a,Eriksen04a, Hansen04, Eriksen05, Eriksen07,deoliveiracosta04,Raeth07, Mcewen08a,Copi08a})
have been made.
These studies have in common that the level of non-Gaussianity 
is assessed  by comparing the results for the measured data with a 
set of simulated CMB-maps which were generated on the 
basis of the standard cosmological model and/or specific assumptions 
about the nature of the non-Gaussianities.\\ 
On the other hand, it is possible to develop model-independent tests 
for higher order correlations (HOCs) by applying the ideas of constrained 
randomisation \cite{Pompilio95a,Raeth02, Raeth03}, 
which have been developed in the field of nonlinear time series analysis 
\cite{Theiler92}.
The basic formalism is to compute statistics sensitive to HOCs
for the original data set and for an ensemble of surrogate data sets,
which mimic the linear properties of the original data.
If the computed measure for the
original data is significantly different from the values obtained
for the set of surrogates, one can infer that the data contain HOCs.\\
Based on these ideas we present in this  {\it Letter} a new 
method for generating surrogates
allowing for probing  scale-dependent non-Gaussianities.
\begin{figure}[h]
\centering
  \includegraphics[width=7.0cm,angle=0]{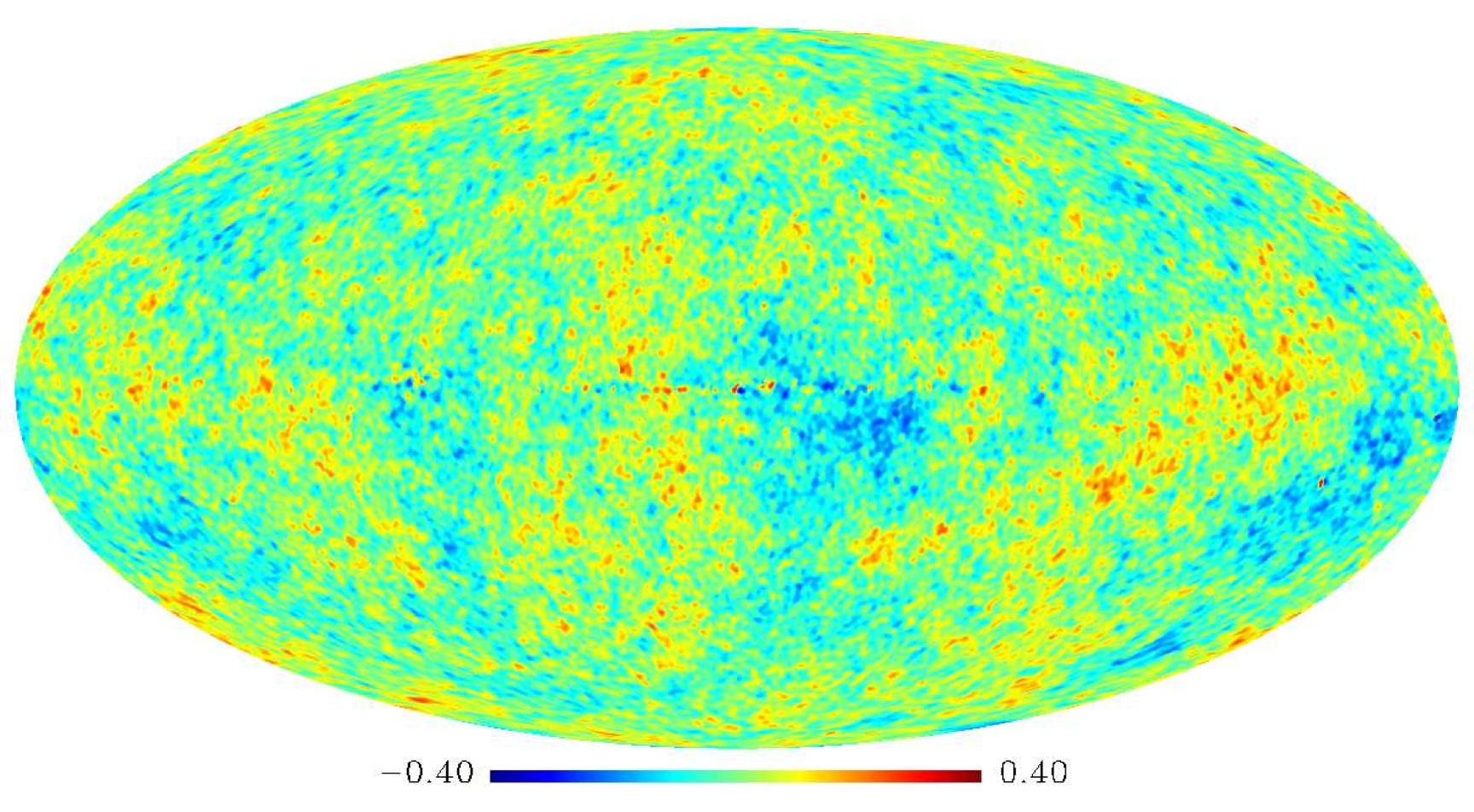}
  \includegraphics[width=7.0cm,angle=0]{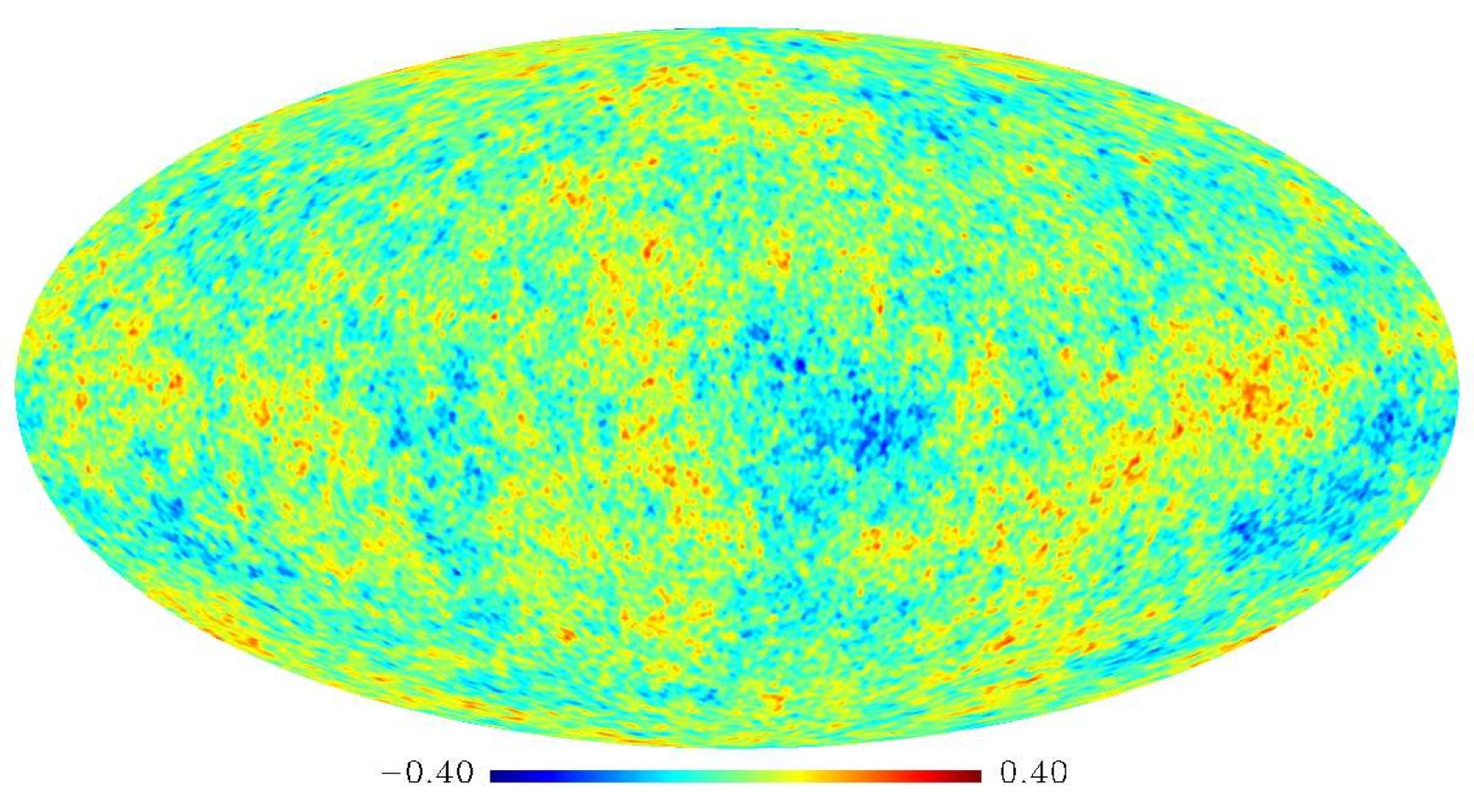} 
   \includegraphics[width=7.0cm,angle=0]{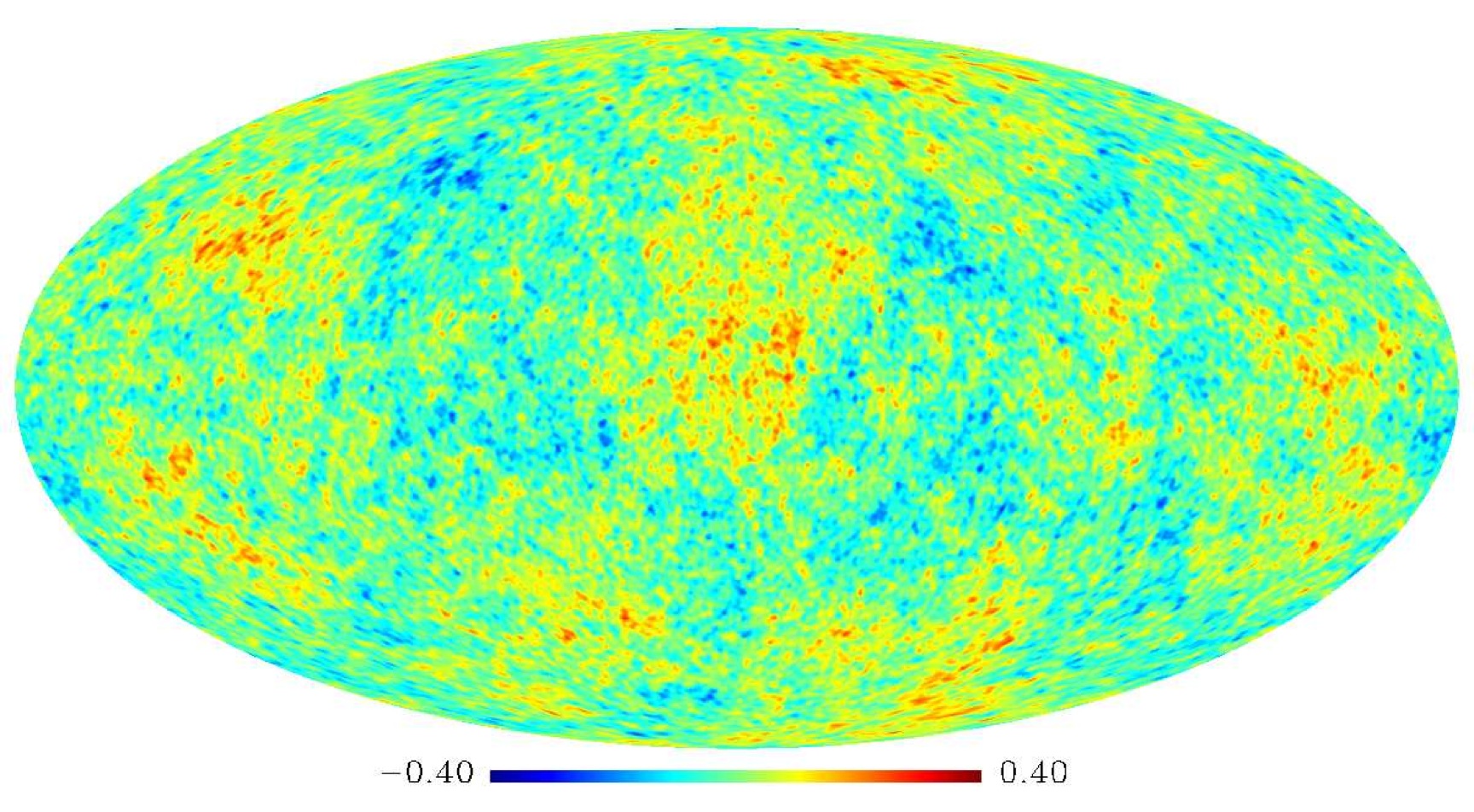}  
\caption{ILC map after remapping of the temperatures and phases (above). 
               First order (middle) and respective second order surrogate (below) for $l_{cut}=20$. 
               Note the resemblance of the first order surrogate with the 
               ILC map at large scales   \label{figure1}}
\end{figure}
Our study is based on the WMAP data of the CMB.
Since our method in its present form requires full sky 
coverage to ensure the orthogonality of the set of
basis functions $Y_{lm}$ we used the 
 five-year  "foreground-cleaned"  
Internal Linear Combination (ILC) map (WMAP5) \cite{Gold09a}
generated and provided\footnote{http://lambda.gsfc.nasa.gov/} by
the WMAP-team.
For comparison we also included the maps produced by Tegmark et al.  \cite{Tegmark03, Deoliveiracosta06},
namely the three year cleaned map (TOHc3) and the 
Wiener-filtered cleaned map (TOHw3)\footnote{http://space.mit.edu/home/tegmark/wmap.html}, which were 
generated pursuing a different approach for foreground cleaning.
Since the Gaussianity of the temperature distribution and the 
randomness of the set of Fourier phases are a necessary 
prerequisite for the application of our method we performed the
following preprocessing steps.
First, the  maps were remapped onto a Gaussian distribution in 
a rank-ordered way. 
By applying this remapping we automatically focus on HOCs induced by the 
spatial correlations in the data while excluding any effects coming from 
deviations of the temperature distribution from a Gaussian one.\\ 
To ensure the randomness of the set of 
Fourier phases we performed a rank-ordered  remapping of the phases 
onto a set of uniformly distributed ones followed
by an inverse Fourier transformation. 
These two preprocessing steps
result in minimal changes to the ILC map 
(the maps remain highly correlated with cross-correlations  $c > 0.95$).
The main effect is the
removal of significant outliers in the temperature distribution.
To test for scale-dependent non-Gaussianities in a model-independent way
we propose the following two-step procedure. Without loss of generality we 
restrict the description of the method and all subsequent analyses 
to the case of non-Gaussianities on large scales.
Consider a CMB map $T(\theta,\phi)$,
where $T(\theta,\phi)$ is Gaussian distributed and calculate its Fourier 
transform. The complex valued Fourier coefficients $a_{lm}$,
$a_{lm} = \int d\Omega_n T(n) Y^{*}_{lm}(n)$
can be written as
$a_{lm} = | a_{lm} | e^{i \phi_{lm}} $ 
with  $\phi_{lm}=\arctan \left( Im(a_{lm}) / Re(a_{lm} )  \right)$.
The linear or Gaussian properties of the underlying random field 
are contained in the absolute values $ | a_{lm} | $, 
whereas all HOCs -- if present -- 
are encoded in the phases $\phi_{lm}$ and the correlations among them.
First, we generate a first order surrogate map, 
in which any phase correlations 
for the scales, which are not of interest (here: the small scales), 
are randomised.  This is achieved by a random shuffle of the phases  $\phi_{lm}$ 
for $l > l_{cut}, 0 < m \le l$, where  $l_{cut}=10, 15, 20, 25, 30$ in this {\it Letter} and
by performing an inverse Fourier transformation (Fig. \ref{figure1}).
Second, $N$ ($N=500$ for $l_{cut}=20$, $N=100$ otherwise) 
realisations of second order surrogate maps 
are generated for the first order surrogate map, in which the remaining 
phases $\phi_{lm}$  with $1 < l \le l_{cut}, 0 < m \le l$ are shuffled while the 
already randomised phases for the small scales are preserved.
Fig. \ref{figure1} shows a realisation  of a second order surrogate map after 
inverse Fourier transformation. Note that the  Gaussian properties  of 
the {\it remapped} ILC map, which are given by $| a_{lm} |$,
are {\it exacly} preserved in all surrogate maps.
Finally, for calculating higher order statistics the maps 
were degraded to $N_{side}=256$ and residual monopole 
and dipole contributions were subtracted. 
\begin{figure}[h]
\centering
  \includegraphics[width=7.5cm,angle=0]{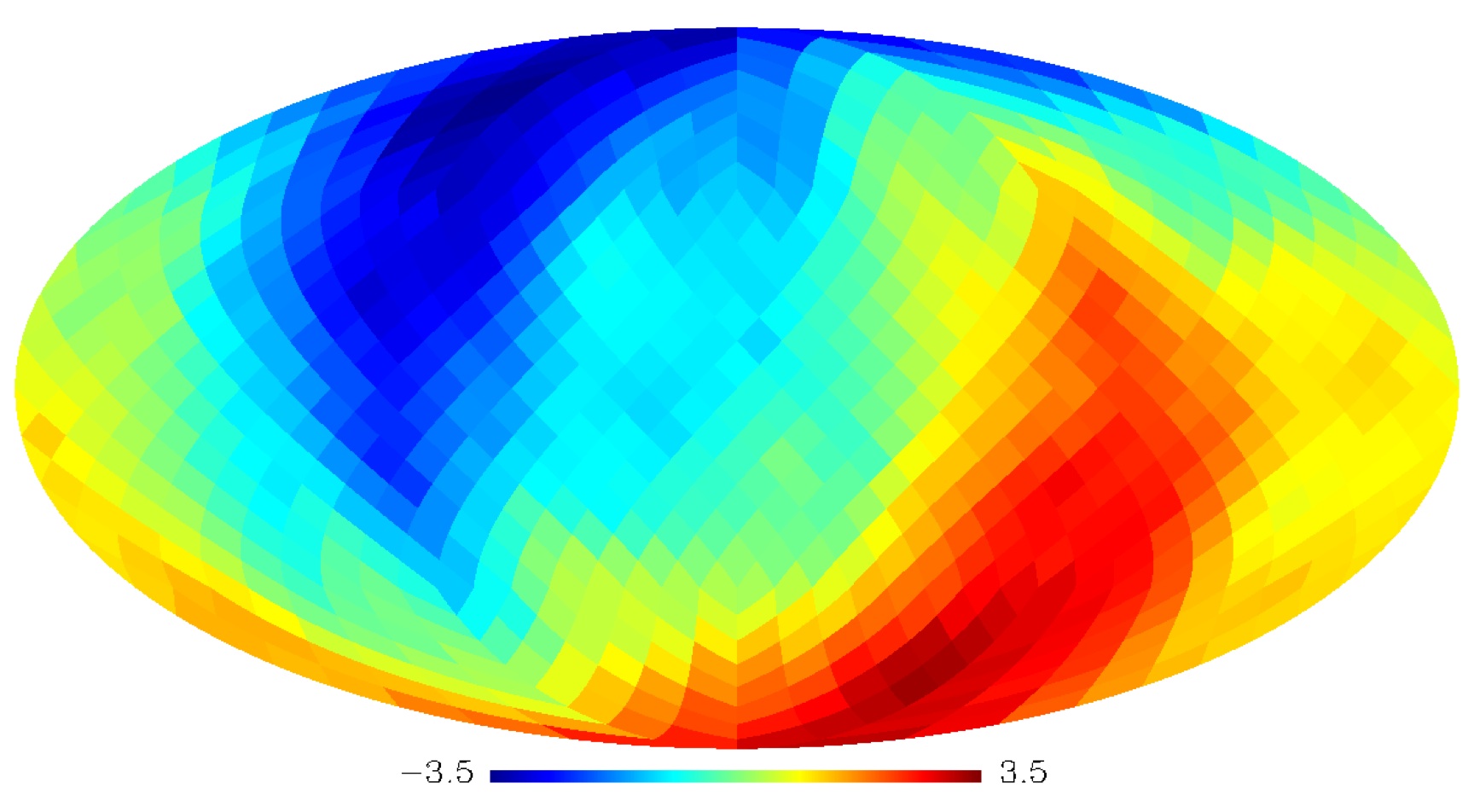}
  \includegraphics[width=7.5cm,angle=0]{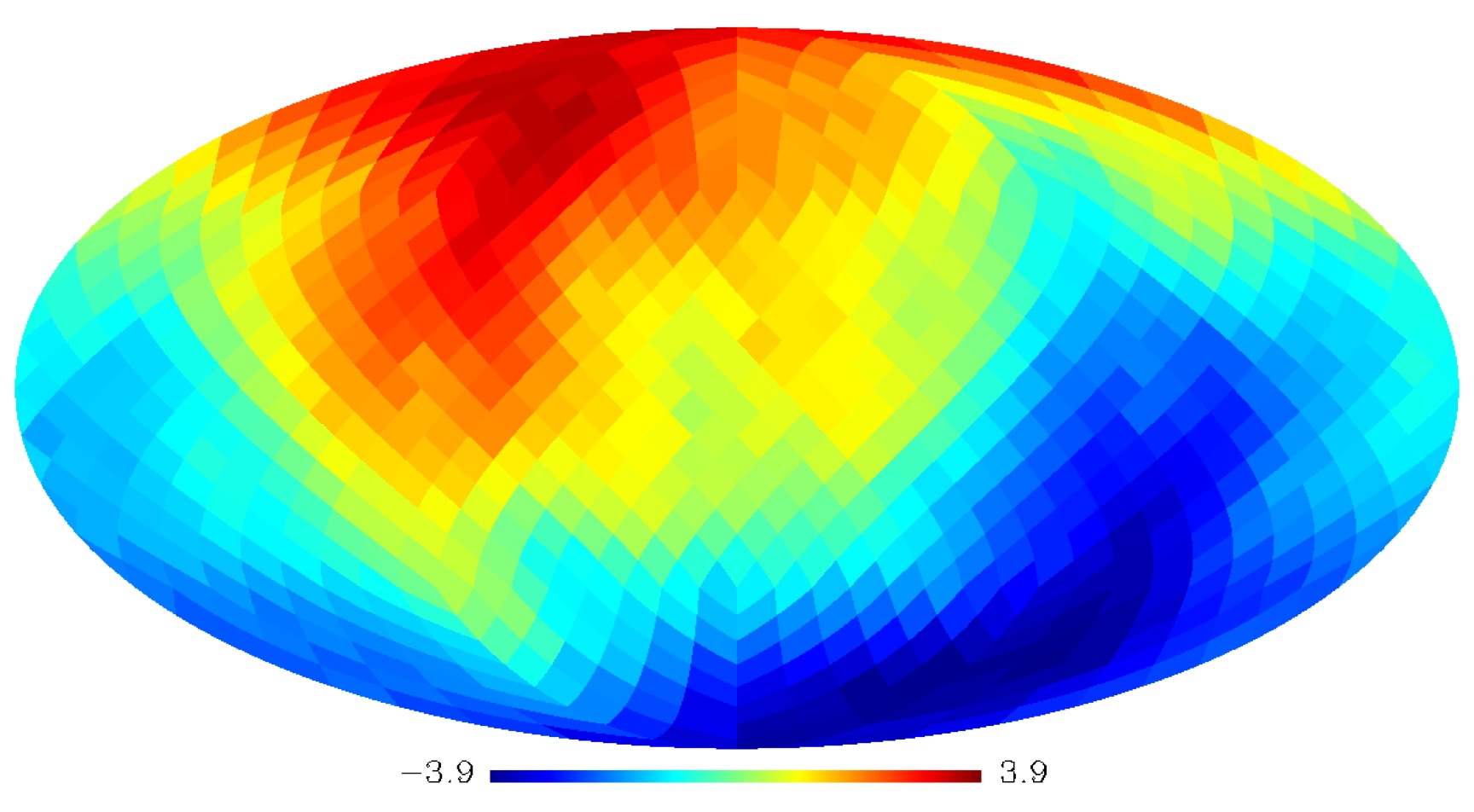}
\caption{Deviation $S$  as derived from rotated upper hemispheres for $\sigma_{T}$ 
                (above) and $\langle\alpha(r_{10})\rangle$ (below) for the WMAP5 map and $l_{cut} =20$. The z-axis 
                of the respective rotated reference frame pierces the centre of the respective 
                colour-coded pixel. $768$ rotated hemispheres, which correspond to number of 
                coloured pixels, were considered. (For a more detailed description of this visualisation 
                technique see e.g. \cite{Hansen04, Raeth07}). \label{figure2}}
\end{figure}
\begin{figure}[h]
\centering
      \includegraphics[width=8.0cm,angle=0]{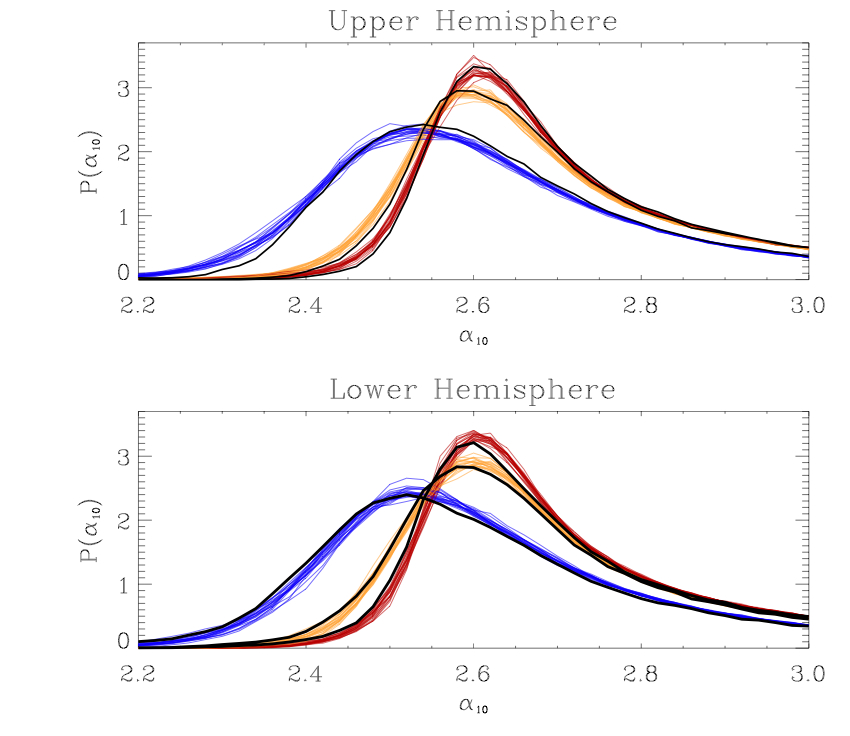}
\caption{Probability density $P(\alpha(r_{10}))$ for the surrogates of the WMAP5 (blue), 
                TOHw3 (yellow) and TOHc3 (red) map for the rotated upper and lower hemisphere and $l_{cut} =20$.
                The black lines denote the respective first order surrogate.  The 
                 reference frame is chosen such that the difference $\Delta S=S_{up} - S_{low}$ 
                 between the upper and lower hemisphere becomes
                 maximal for $\langle\alpha(r_{10})\rangle $ regarding the WMAP5 surrogates.  \label{figure3}  } 
\end{figure}
To compare the two classes of surrogates, we calculate local statistics
in the spatial domain, namely scaling indices (SIM)  
as described in  R\"ath et al. \cite{Raeth07}.
In brief, scaling indices estimate
local scaling properties of a point set $P$.
The spherical CMB data can be represented as a
three-dimensional point distribution $P=\vec{p_i} = (x_i,y_i,z_i), i=1,\ldots,N_{pixels}$ 
by transforming the temperature fluctuations into a radial jitter. 
For each point $\vec{p_i}$ the local weighted cumulative point distribution
$\rho$ is calculated  $\rho(\vec{p_i},r) = \sum_{j=1}^{N_{pixels}} e^{-(\frac{d_{ij}}{r})^2} \;, 
d_{ij} = \| \vec{p_i} - \vec{p_j} \| $.
The weighted scaling indices $\alpha(\vec{p_i},r)$ are then obtained by calculating
the logarithmic derivative of $\rho(\vec{p_i},r)$ with respect to $r$,
 $\alpha(\vec{p_i},r) = \frac{\partial \log \rho(\vec{p_i},r)}{\partial \log r}$. 
For each pixel we calculated scaling indices for ten different 
scales, $r_1=0.025$,\ldots,$r_{10}=0.25$ in the notation of \cite{Raeth07}.
For each scale we calculate the mean ($\langle \alpha \rangle$) and standard deviation ($\sigma_{\alpha}$)  
of the scaling indices  $\alpha(\vec{p_i},r)$ derived from a set of pixels belonging 
to rotated hemispheres or the full sky. 
To investigate the correlations between the scaling indices and 
temperature fluctuations, we also considered the standard deviation ($\sigma_{T}$)  for 
the mere temperature distribution of the respective sky regions. 
\begin{figure}[h]
\centering
  \includegraphics[width=8.0cm,angle=0]{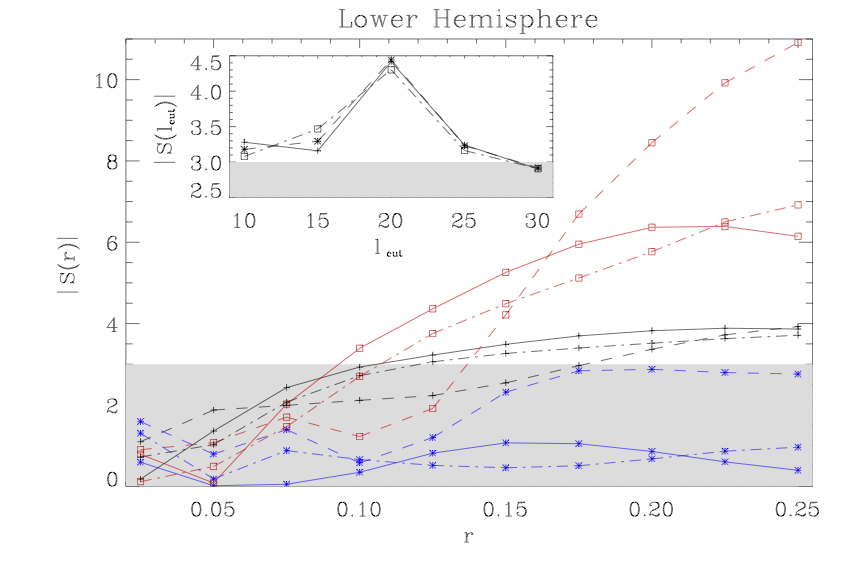}
  \includegraphics[width=8.0cm,angle=0]{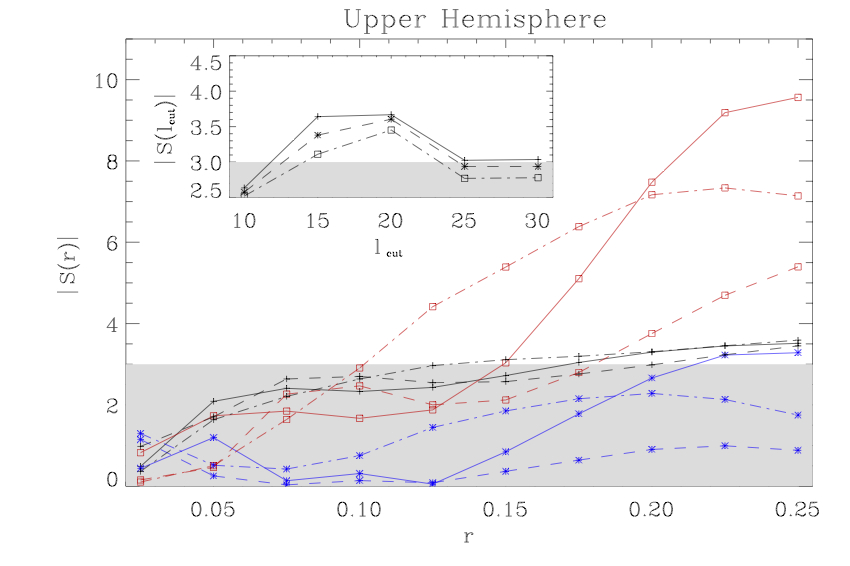}
\caption{Deviations $|S(r)|$ for the rotated upper and lower hemisphere 
                for $\langle\alpha\rangle$ (black),
                $\sigma_{\alpha}$ (blue) and a  $\chi^2$-combination of  $\langle \alpha \rangle$ 
                and  $\sigma_{\alpha}$ (red) ($l_{cut} =20$, $N=500$).
                The solid (dashed, dashed-dotted) lines denote the WMAP5 (TOHw3, TOHc3) map.  The shaded region
                 indicates the $3 \sigma$ significance interval.  The insets show the results for $\langle \alpha(r_{10}) \rangle$,
                 $\langle \alpha(r_{9}) \rangle$ and $\langle \alpha (r_{8})\rangle$ (solid, dashed, dashed-dotted)
                 as a function of $l_{cut}$ for the WMAP5 map (here: $N=100$). \label{figure4} }
\end{figure}
\begin{table} 
\caption{S/SL Upper Hemisphere \label{table1}}
\begin{ruledtabular}
\begin{tabular}{lccc}
     &  WMAP5  & TOHc3 & TOHw3   \\
     &  ($S$/$SL$) & ($S$/$SL$) & ($S$/$SL$)\\ \hline
$\sigma_{T}$ & -2.8/ 99.8 & -3.0/$>$99.8 & -2.9/ $\>$99.8  \\
$\langle \alpha (r_{10})\rangle$ & 3.5 / $>$99.8 & 3.5 /  $>$99.8 & 3.6 /  $>$99.8 \\
$\chi^2_{\langle \alpha \rangle}$ & 5.7 / 99.8 & 5.2 /  99.6 & 7.0 /  $>$99.8\\
$\chi^2_{\sigma_{\alpha}}$ & 3.1 / 99.2 & -0.7 / 74.4 &  2.1 / 95.4\\
$\chi^2_{\langle \alpha \rangle, \sigma_{\alpha}}$ & 6.1 / $>$ 99.8 & 3.6 / 99.0  & 6.4 / $>$ 99.8\\
\end{tabular}
\end{ruledtabular}
\end{table}
\begin{table} 
\caption{S/SL Lower Hemisphere \label{table2}}
\begin{ruledtabular}
\begin{tabular}{lccc}
     &  WMAP5  & TOHc3 & TOHw3   \\
     &  ($S$/$SL$) & ($S$/$SL$) & ($S$/$SL$)\\ \hline
$\sigma_{T}$ & 2.7/99.8 & 2.9/$>$99.8 & 2.8/99.8 \\
$\langle\alpha (r_{10})\rangle$ & -3.9 / $>$99.8 & -3.9 / $>$99.8 & -3.7/$>$99.8 \\
$\chi^2_{\langle\alpha\rangle}$ & 7.9 /$>$99.8 & 5.4/99.8 & 7.3/$>$99.8\\
$\chi^2_{\sigma_{\alpha}}$ & -0.7 / 76.4 & 4.4 / 99.6 &  -0.6/67.0\\
$\chi^2_{\langle\alpha\rangle, \sigma_{\alpha}}$ & 5.8 / 99.8 & 6.3 /$>$99.8 & 5.2/$>$99.8   \\
\end{tabular}
\end{ruledtabular}
\end{table}
The differences of the two classes of surrogates 
are quantified by the $\sigma$-normalised deviation 
$S(Y)= (Y_{surro1} - \langle Y_{surro2} \rangle)/ \sigma_{Y_{surro2}}$, 
$Y=\sigma_{T}$,$\langle \alpha \rangle$,$\sigma_{\alpha}$,$\chi^2$  
(surro1: first order surrogate, surro2: second order surrogate) and
the significance levels $SL = 1 - p$, where $p$ is the fraction 
of second order surrogates, which have a higher 
(lower) $Y$ than the first order surrogate.
$\chi^2$ denotes diagonal  $\chi^2$-statistics, which we obtain 
by combining $\langle \alpha \rangle$,$\sigma_{\alpha}$ for a given scale $r_i$, i.e.
$\chi^2(r_i) =   \sum_{j=1}^{2}  \left[ \frac{X_j(r_i) - \langle X_j(r_i)\rangle}
                           {\sigma_{X_j(r_i)}} \right]^2 \;,$ 
                 with $X_1=\langle \alpha \rangle, X_2= \sigma_{\alpha} $ and 
                 $\langle X_{j}  \rangle$,$\sigma_{X_{j} }$ derived from 
the $N$ realisations of  second order surrogates.
As scale-independent measure we also consider  $\chi^2$ as obtained 
by summing over the scales ($N_r=10$), 
$\chi^2 =  \sum_{i=1}^{N_r} \sum_{j=j_1}^{j_2}  \left[ \frac{X_j(r_i) - \langle X_j(r_i)\rangle}
                 {\sigma_{X_j(r_i)}} \right]^2 \;,$
for one single  measure  
($j_1=1, j_2 =1$; $j_1=2, j_2 =2$) and  the two measures ($j_1=1, j_2 =2$).
Fig. \ref{figure2} shows $S(\sigma_{T})$ and $S(\langle \alpha(r_{10}) \rangle)$ derived from 
pixels belonging to  the respective upper hemispheres for $768$ rotated reference frames.
Statistically significant signatures
for non-Gaussianity and ecliptic hemispherical asymmetries 
become immediately obvious, whereby these signatures can solely be induced by large scale HOCs.  
Although  $S(\sigma_{T})$ and $S(\langle \alpha(r_{10}) \rangle)$ are spatially 
highly (anti-)correlated ($c=-0.95$), the two effects are nevertheless complementary 
to each other in the sense that a systematically 
lower/higher $\sigma_{T}$  would lead  to a lower/higher $\langle \alpha(r_{10}) \rangle$ and not
to the observed higher/lower value for the first order surrogate map.
These systematically shifted scaling indices are a generic feature
present in all three maps (Fig. \ref{figure3}). Although the probability densities 
$P(\alpha(r_{10}))$ are different due to the smoothing or Wiener-filtering 
for the three maps, the shifts of the first order
surrogate relative to its second order surrogates can be found 
in all three cases. We also cross-correlated  the deviation maps shown in Fig. \ref{figure2}
derived from the three input maps and always obtained $c \ge 0.98$ for the correlation
coefficient.
These systematic deviations lead to significant 
detections of non-Gaussianities which are shown in Fig. \ref{figure4} and
summarised for $l_{cut} =20$ in Tables \ref{table1}-\ref{table2}.
The most significant and most stable results are found for $\langle\alpha\rangle$ 
at larger radii, where for all three maps none of the $500$ second order 
surrogates had a  higher (upper hemisphere) or lower (lower hemisphere) value
than the respective first order surrogate, leading to a significance 
level $SL > 99.8$ \% for $\langle\alpha(r_{10})\rangle$.
Also the combined measure $\chi_{\langle\alpha\rangle}^2$ yields  deviations $S$ 
ranging from $5.2$ up to $7.9$,  which represent one of the most significant 
detection of non-Gaussianity in the WMAP data to date. 
We estimated how varying $l_{cut}$ values affect the results and found that both
the non-Gaussianities and asymmetries are detected for all considered $l_{cut}$, 
where the highest deviations are obtained for $l_{cut} = 20$. Although $S$ becomes
considerably smaller for $l_{cut} = 10$, we can still detect the non-Gaussianities
with $SL > 99.0$ \%, which is larger than the results reported 
in \cite{Chiang07} ($SL = 95$ \%), where also $l_{cut} = 10$ was used.
We perfomed the same 
analyses for the coadded WMAP foreground template maps and for 
simulations using the best fit $\Lambda$CDM
power spectrum and WMAP-like noise and beam properties.
We found in none of these cases significant 
signatures as reported above. Details about these studies 
are deferred to a longer forthcoming publication.\\
In conclusion, we demonstrated  the feasibility to generate
new classes of surrogate data sets preserving the power spectrum and partly
the information contained in the Fourier phases, while all other 
HOCs are randomised.
We found significant evidence for both asymmetries and 
non-Gaussianities on large scales in the WMAP data of the CMB 
using scaling indices as test statistics.
The novel statistical test involving new classes of surrogates
allows for an unambigous relation of the signatures 
identified in real space with scale-dependent HOCs, which 
are encoded in the respective Fourier phase correlations.
Our results, which are consistent with previous 
findings \cite{Park04a,Eriksen04a,Hansen04,Eriksen05,Eriksen07,Chiang07,Raeth07} but also extend
to smaller scales than those reported in \cite{deoliveiracosta04} ($l_{cut} = 3$) ,
\cite{Chiang07}  $(l_{cut} = 10)$ and \cite{Copi08a} 
($l_{cut} \le 3$), 
point towards a violation of statistical isotropy and Gaussianity.
Such features would disfavour 
canonical single-field slow-roll inflation -- unless there is some undiscovered systematic 
error in the collection or reduction of the CMB data or yet unknown foreground contributions.
Thus, at this stage it is too early to claim the detected HOCs as
cosmological and further tests are  required to elucidate the true origin of the 
detected anomalies. Their existence in the three maps might, however, be suggestive.\\
In either case the proposed statistical method  offers an efficient  tool
to develop model-independent tests for scale-dependent  non-Gaussianities.
Due to the generality of this technique it can be applied to any signal,
for which the analysis of scale-dependent HOCs is of interest.\\
Many of the results in this paper have been obtained using 
HEALPix  \cite{Gorski05}. 
We acknowledge the use of LAMBDA. Support for LAMBDA is provided by the 
NASA Office of Space Science.

\end{document}